\documentclass{article}
\usepackage{PRIMEarxiv}
\usepackage{amsmath}
\usepackage[T1]{fontenc}    
\usepackage{hyperref}       
\usepackage{url}            
\usepackage{booktabs}       
\usepackage{amsfonts}       
\usepackage{nicefrac}       
\usepackage{microtype}      
\usepackage{lipsum}
\usepackage{fancyhdr}       
\usepackage{graphicx}       
\usepackage{xcolor}

\title{Geographical Isolation as a Driver of Political Violence in African Cities}

\author{
  Rafael Prieto-Curiel\\
  Complexity Science Hub \\ Josefstaedter Strasse 39 \\ 1080 Vienna, Austria \\
  \texttt{prieto-curiel@csh.ac.at} \\
   \And
  Ronaldo Menezes \\
  BioComplex Laboratory, Computer Science\\ University of Exeter, EX4 4QJ, Exeter, UK \\ \\
  Computer Science \\ Federal University of Cear\'a, \\ 60020-181, Fortaleza, Brazil
}

\begin{document}
\maketitle

\section{Abstract}

Violence is commonly linked with large urban areas, and as a social phenomenon, it is presumed to scale super-linearly with population size. This study explores the hypothesis that smaller, isolated cities in Africa may experience a heightened intensity of violence against civilians. It aims to investigate the correlation between the risk of experiencing violence with a city's size and its geographical isolation. Over a 20-year period, the incidence of civilian casualties has been analysed to assess lethality in relation to varying degrees of isolation and city sizes. African cities are categorised by isolation (number of highway connections) and centrality (the estimated frequency of journeys). Findings suggest that violence against civilians exhibits a sub-linear pattern, with larger cities witnessing fewer casualties per 100,000 inhabitants. Remarkably, individuals in isolated cities face a quadrupled risk of a casualty compared to those in more connected cities.

\section{Introduction}

{
Violence is a pressing global issue that causes nearly half a million deaths worldwide each year \cite{WhoHomicides, UNODCHomicides}. Although different parts of the world suffer distinct expressions of violence, it is the main reason for concern in many countries. For example, cartels and organised crime in Latin America \cite{PrietoCampedelliHope2023}, politically motivated groups in Africa \cite{radil2022urban, OWReportOECD}, and mass shootings in the USA \cite{schildkraut2021framing} are some of the most severe manifestations of violence today. Violence has displaced millions, often across international borders \cite{MigrationConflictColombia, ConflictMigrationSaharanAfrica, dreher2011hit}. Violence imposes a massive burden on the quality of life, creates lifetime uncertainties, deters investment, and significantly impacts the economy, amounting to nearly 10.9\% of global GDP \cite{enders1996terrorism, eckstein2004macroeconomic, IEP2022, sloboda2003assessing, index2023mexico, aburto2023global}. Estimates suggest that in some countries, like Syria, Afghanistan, and Iraq, the economic impact of violence exceeds 50\% of GDP \cite{iqbal2021estimating}. Despite differing causes and forms, these violent phenomena share commonalities. They are not random but occur under conditions where targets, perpetrators, and enabling factors (like weak state governance) coincide \cite{RoutineActivity, CriminalityOfPlace}. These scenarios often repeat, suggesting that one event increases the likelihood of observing subsequent incidents nearby, thus forming spatial and temporal patterns crucial for predicting future violence \cite{chuang2019mathematical, porter2012self, siebeneck2009spatial, SystematicReviewPlaces, CrimeConcentrationVaryingCitySize, oliveira2017scaling, bahgat2013overview, ge2022modelling}.
}

{
Although rural and sparsely-populated areas might be perceived as conducive to insurgencies due to limited state reach, they rarely offer a rich set of targets for violent activities. Violence mainly manifests as an urban phenomenon \cite{radil2022urban, LawCrimeConcentration, prieto2021heartbeat}. Cities aggregate populations, potential targets and victims, resources for armed factions, intra-elite conflicts, and individuals susceptible to involvement in violence or recruitment by violent groups \cite{menashe2020migrant, urdal2012explaining, OWReportOECD}. The coexistence of extreme wealth and poverty is a characteristic urban feature \cite{OWReportOECD}. Urban environments serve as battlegrounds for political, economic, and religious elites, symbolising state authority \cite{beall2013cities, staniland2010cities}. The urban setting facilitates criminals blending with the rest of the population, thus lowering capture risks. Additionally, cities, with their enhanced media presence and infrastructure, become focal points for insurgents seeking public and media attention \cite{zhukov2012roads, PrietoCEUSMedia}. Notably, most terrorist and conflict incidents occur in strategic locations, including densely populated regions, military zones, or city centres \cite{savitch2014cities, raleigh2010seeing}.
}

{
Yet, although violence predominantly occurs in urban settings, not all cities experience violence to the same extent. Significant heterogeneity in violence levels exists across cities, even within the same country, or among cities with similar characteristics \cite{SacerdoteCrimeInteractions}. It has been previously argued that city size plays a crucial role in influencing violence levels \cite{UrbanScalingWest, GrowthBettencourt}. In some cases, it has been observed that violent crime rates per capita are higher in larger cities compared to smaller towns or rural areas \cite{SacerdoteCrimeCities, CrimeAndUrbanFlight, GrowthBettencourt}. For instance, larger Brazilian cities report more homicides per 100,000 inhabitants than smaller ones, with similar trends observed for homicides in Colombia, burglaries in Denmark, and property crime in the USA \cite{alves2013scaling, gomez2012statistics, oliveira2021more, chang2019larger}. However, opposite indicators have also been observed, where larger cities are less violent, such as murders and homicides in India and burglary in South Africa or Canada \cite{sahasranaman2019urban, oliveira2021more}. Similarly, it has been detected that the death toll and the number of terrorist events are uncorrelated to city size \cite{guo2019common}. Thus, city size has been wrongfully assumed to capture violence heterogeneities, and large cities have also been wrongfully deemed to be more violent than small cities. City size has a limited impact on the levels of violence experienced at the personal level, and the effect is country-specific and depends on the type of violence as well. The reality is that other aspects, such as physical geography, proximity, and resources, have more influence on violent patterns than a mere look at size \cite{raleigh2010seeing, buhaug2008contagion}.
}

{
Here, the impact of isolation on violence ---beyond looking only at city size--- is analysed. Some places are more connected, accessible and central, whilst others are remote, vulnerable and secluded \cite{van2021being, latora2005vulnerability, walther2020mapping, prieto2022detecting, ganin2017resilience, liang2024intercity}. It has been suggested that secluded areas rarely offer targets for attacks, and therefore, insurgencies are drawn into more central areas \cite{buhaug2008contagion, zhukov2012roads}. Similarly, places governing access to other sites are more valuable targets for groups fighting over power and thus could be frequently fought over \cite{hammond2018maps}. However, an opposite theory suggests that a weaker state results in a higher probability of violence, meaning that remote locations (with low state capacity) have an increased risk of armed conflict \cite{fearon2003ethnicity, muller2021roads}. Thus, detecting whether central cities or remote and low-accessibility towns have more violence is an open question. 
}

{
This study examines if isolated cities experience higher levels of violence and casualties compared to more central and well-connected cities, with a focus on politically-motivated violence. Africa's murder rate per 100,000 population is four times higher than in Europe, showing a significant increase over the past decade. By analysing the proximity of violent events to urban areas, this research aims to determine the influence of city size, centrality, and isolation on violence frequency. By leveraging data from the Armed Conflict Location \& Event Data project (ACLED) \cite{raleigh2010introducing}, the study employs two urban indicators to assess a city's isolation level: the number of connecting roads and a proxy for the frequency of intracity journeys. The findings underscore the significant impact of geographical isolation on violence. Specifically, individuals residing in isolated cities encounter a markedly higher rate of casualties --four times greater-- compared to those in centrally located cities. This highlights isolation not merely as a contributing factor but as a critical determinant in the prevalence and severity of violence experienced by urban populations.
}

\section{Results}

{
The correlation between city size and violence is assessed by fitting the equation \(V_i = \alpha P_i^\beta\), where \(V_i\) is some metric of violence (for example, the number of events or casualties), \(P_i\) is the population of the city \(i\), and where \(\alpha\) and \(\beta\) are the model parameters. Values of \(\beta > 1\) suggest that people in larger cities suffer more violence. It is frequently assumed that large cities concentrate more violence than small cities. This assumption is often called ``universal'', often with a fixed scaling coefficient of approximately 1.15, and is frequently explained as the result of the increased number of interactions or social activity or increasing costs for providing security in big cities \cite{ScalingInteractions, GrowthBettencourt, CrimeAndUrbanFlight, SacerdoteCrimeCities, UrbanScalingWest}. For example, it has been found that \(\beta = 1.26\) in the case of theft in Mexico, and \(\beta = 1.16\) in the case of serious crime in the US \cite{oliveira2021more, GrowthBettencourt}, among others (see Figure \ref{WorldBeta} for an extensive list of all known studies of the scaling of violence). However, a few notable exceptions have been found, such as burglary in South Africa (\(\beta = 0.91\)) \cite{oliveira2021more}, and the number of crimes (\(\beta = 0.8699\)) and homicides (\(\beta = 0.7788\)) in India \cite{sahasranaman2019urban}. For an extensive list of scaling coefficients related to violence, refer to the SM-C.

\begin{figure}[ht]
  \centering
    \includegraphics[width=.8\textwidth]{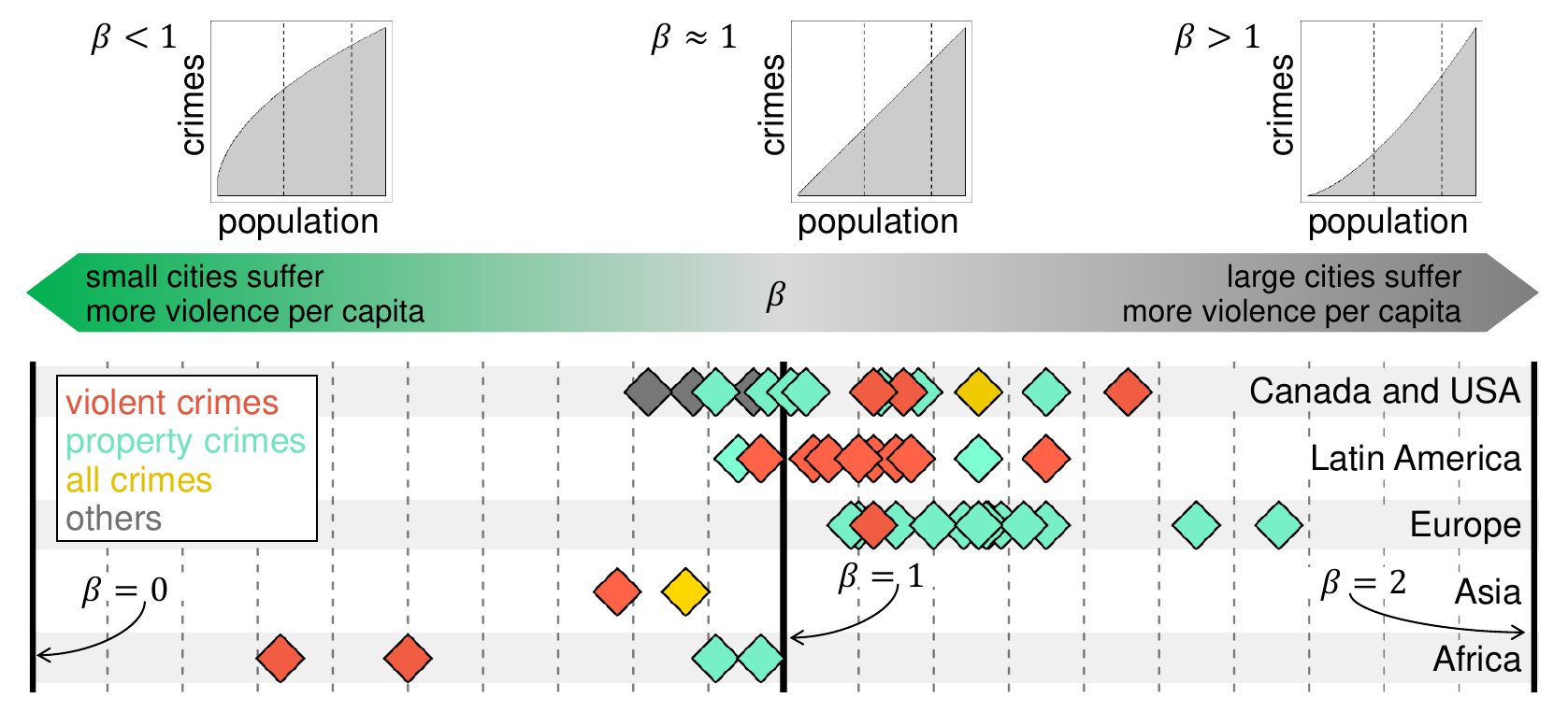}
    \caption{\textbf{Scaling Coefficients Across the World.} Observed scaling coefficients for different types of crime and violence suffered in some parts of the world.  Coefficients obtained for Brazil \cite{ignazzi2014scaling}, Canada \cite{oliveira2021more}, Colombia \cite{gomez2012statistics}, England and Wales \cite{hanley2016rural}, Italy \cite{oliveira2021more}, Mexico \cite{oliveira2021more}, USA \cite{banerjee2015competitive, GrowthBettencourt, gomez2016explaining} and the UK \cite{oliveira2021more}. More coefficients are in the SM-C.
    }
    \label{WorldBeta}
\end{figure}
}

{
To quantify the correlation between city size and violence in Africa, we use two datasets. The ACLED dataset tracks political violence globally through local media reports. In Africa, nearly 300,000 politically-motivated events associated with nearly 600,000 casualties were reported between January 2000 and October 2022 \cite{raleigh2010introducing}. Events are classified into six non-overlapping categories: 25\% are classified as battles (interactions between two organised armed groups), 25\% as violence against civilians (where an organised armed group deliberately inflicts violence upon unarmed non-combatants), 24\% as protests, 11\% as riots, 8\% as explosions, and 7\% as strategic developments (which include non-violent activity by conflict actors, such as arrests). We then consider the delineation of urban agglomerations from Africapolis and, for each city, count the number of reported events and casualties within each polygon \cite{Africapolis}. We find that the number of events, casualties, attacks against civilians, and casualties they have created are all sublinear. For events reported since 2015, the number of events has a scaling coefficient $\beta_E \approx \beta_L \approx 2/3$, and similarly for violence against civilians (all coefficients by year are in the SM-A). Thus, looking directly at the events which are recorded within each urban polygon in Africa, we detect a sublinear correlation with city size. People in smaller African cities are more violent and suffer more violence against civilians than people in larger cities.
}

{
Although it is possible to use the boundary of each urban polygon, that process creates a classification based on two datasets that are not highly accurate at the spatial level. On the one hand, ACLED events are often recorded in strategic locations and often repeated. For example, over 1000 events registered in Nouakchott (nearly 99\% of them) have been recorded with the exact same location, which corresponds to some residential building, with no particular relevance. On the other hand, the delineation of Africapolis was constructed based on aerial pictures, often taken many years ago, which would not capture the expansion of African cities. Thus, events that are located maybe a few metres away are not considered part of a city (more details regarding these spatial issues in the SM-A). Instead, here we measured the distance between the event's location and the centre of the nearest city. Events are assigned to the nearest city if they occur within a distance \(\delta\) of the centre of the city. Then, the number of violent events and deaths are counted for each city. With this technique, we detect that half of the violence against civilians and 47\% of the casualties occurred within \(\delta = 20\)km of the centre of an African city (details in the SM-B). Yet, Africa is a vast territory, where less than 8\% of its surface is within 20km of the centre of a city. Thus, violence is mostly urban (Figure \ref{MapEvents}). Further, results show that urban violence is highly heterogeneous and concentrated in a few cities (as observed with violent crime and other events elsewhere \cite{CrimeConcentrationHistory, SystematicReviewPlaces}). The 5\% most violent cities in Africa had 73\% of the fatalities related to the events in the continent since 2000. In contrast, nearly two-thirds of African cities have less than one casualty registered yearly related to this type of event. Yet, the high concentration of events and fatalities across African cities does not correspond to the concentration of the population. Our results show that in the 5\% most violent cities (with 73\% of the continental fatalities), only 15\% of Africa's urban population is currently living.

\begin{figure}[ht]
  \centering
    \includegraphics[width=.8\textwidth]{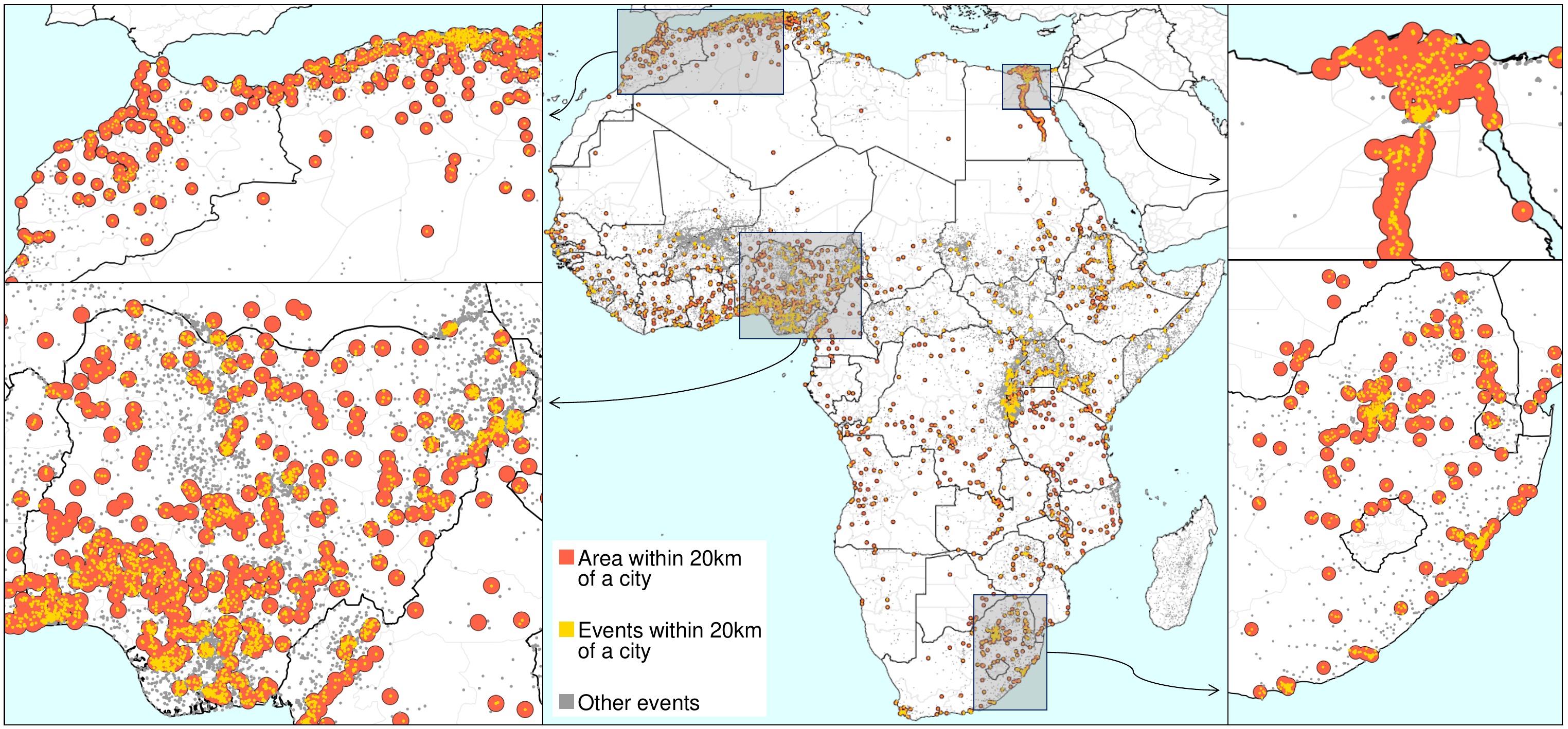}
    \caption{\textbf{Urban Centres and Event Proximity in Africa.} Most events in Africa occur within a short distance of city centres. This study maps the locations of the largest 2,100 cities on the continent, based on data from \cite{Africapolis}, and identifies all reported events (in yellow) within a 20km radius of any urban centre. Events depicted in grey are situated further away from urban centres. The areas within 20km of a city centre are marked in red, illustrating the spatial distribution of events in proximity to urban locales.}
    \label{MapEvents}
\end{figure}
}

{
Violence against civilians is markedly more lethal in smaller African cities. The scaling coefficients for the number of casualties, $\beta_L(\delta)$, and the fatalities related to violence, $\beta_V(\delta)$, are both below 1 -- in fact, below 0.5 for all values of $\delta$ (details in Section \ref{sec:methods}). For instance, it is observed that $\beta_L(20) = 0.4865 \pm 0.0013$ and $\beta_V(20) = 0.3404 \pm 0.0025$, indicating that larger cities report fewer casualties per 100,000 inhabitants compared to smaller cities. This trend persists across different $\delta$ values. For instance, for $\delta = 10$ we get that $\beta_L(10) = 0.4954 \pm 0.0014$ and $\beta_V(10) = 0.33349 \pm 0.0029$. Despite a higher absolute number of casualties in larger cities, when adjusted for population size, violence against civilians is significantly more lethal in smaller cities (details in the SM-C).
}

{
Violence has become increasingly lethal in smaller African cities, particularly in recent years. Analysing casualties from the year 2000, the scaling coefficient stands at $\beta_L = 0.4948$, while for violence against civilians, it is $\beta_V = 0.3369$. However, for events occurring after 2015, these coefficients shift to $\beta_L = 0.3431$ and $\beta_V = 0.1739$, respectively (Figure \ref{ResultsFigure}). Restricting the analysis to cover only events between 2020 and 2022 reveals $\beta_L = 0.1851$ and $\beta_V = 0.0308$, suggesting an increasing trend of violence moving away from larger cities (details in the SM-D). We find that violence in Africa is sublinear, meaning that small cities have more violence per capita than large cities. Thus, the ``universality'' of scaling laws for crime (as discussed by Bettencourt et al. \cite{GrowthBettencourt}) applies only to certain countries and types of violence, but not in Africa for violence against civilians. 

\begin{figure}[ht]
  \centering\includegraphics[width=.8\textwidth]{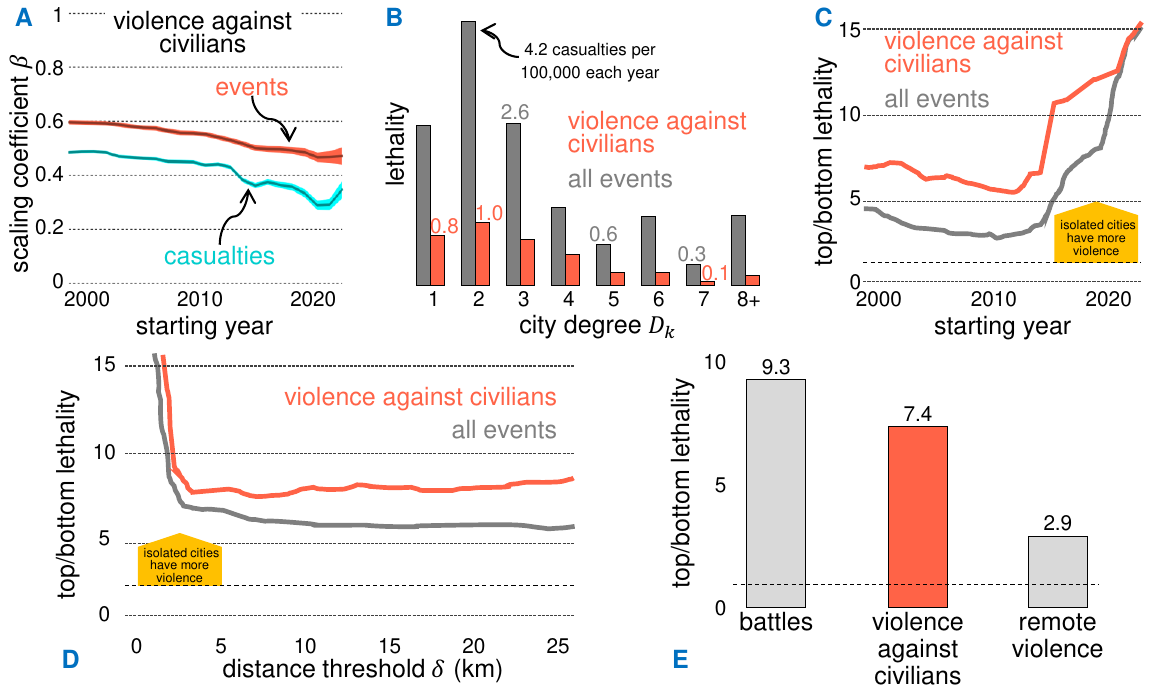}
    \caption{\textbf{Analysis of Urban Lethality Dynamics.} (A) scaling coefficient of violence against civilians suffered across cities of varying size by taking the events starting from a varying year. (B) Lethality (vertical axis) across cities of varying connectivity (horizontal axis). (C) Disparity in lethality (vertical axis) over time, considering events from specific years (horizontal axis). (D) Comparison of highest vs. lowest lethality (vertical axis) based on the distance threshold $\delta$ (horizontal axis) applied to attribute events to cities. (E) Lethality contrast (vertical axis) by type of event, merging 'protests and riots' as well as 'battles and strategic developments' into unified categories from the ACLED database. The dashed line indicates a lethality rate of $\phi = 1$, the hypothetical rate if the most isolated cities experienced the same lethality as the most central cities.}
    \label{ResultsFigure}
\end{figure}
}

\subsection{Violence is more prominent in isolated cities}

{
The 10\% most populous cities of Africa have 66\% of the population but only 33\% of the fatalities related to politically-motivated violence over the past 22 years. Thus, big cities are not inherently violent, particularly compared to other cities that concentrate higher levels of violence: isolated cities. City connectivity explains the emergence of dominant cities and their economic development and innovation patterns \cite{liang2024intercity, pitts1965graph, nystuen1961graph, tao2020urban, castells2011rise, alves2021commuting}. Isolation is one of the main contributors to poverty and a violence generator \cite{LinardPopulationPatterns}. Intercity connectivity has been used to improve scaling models \cite{liang2024intercity}. In China, for example, the number of patents in a city is better explained by a city's mobility network than its size \cite{liang2024intercity}. Data corresponding to the location of major highways in Africa from OpenStreetMap was used to measure their level of isolation \cite{OpenStreetMap, prieto2022constructing}. The \emph{degree} of city $j$, expressed as $D_j$, is the number of highways that connect that city to others \cite{prieto2022detecting}. We classify cities into three groups depending on their degree, with roughly the same population in each group (Figure \ref{FigureDiagramMS}). Cities are labelled as ``high isolation'' if $D_j \leq 2$, ``medium isolation'' if $D_j \in [3,5]$, and ``low isolation'' if $D_i \geq 6$. Although most cities are highly isolated, they tend to be small. In cities characterised by high isolation, the rate of registered casualties stands at 3.9 per 100,000 people annually, in stark contrast to just 0.7 per 100,000 in cities with low isolation levels. This stark discrepancy underscores that lethality in highly isolated cities is 5.4 times higher than in their less isolated counterparts (Figure \ref{ResultsFigure}-A). Specifically, when considering violence against civilians, the lethality rate in cities with high isolation escalates to 7.4 times that of more centrally located cities.

\begin{figure}[ht]
\begin{center}
\includegraphics[width = 0.9\linewidth]{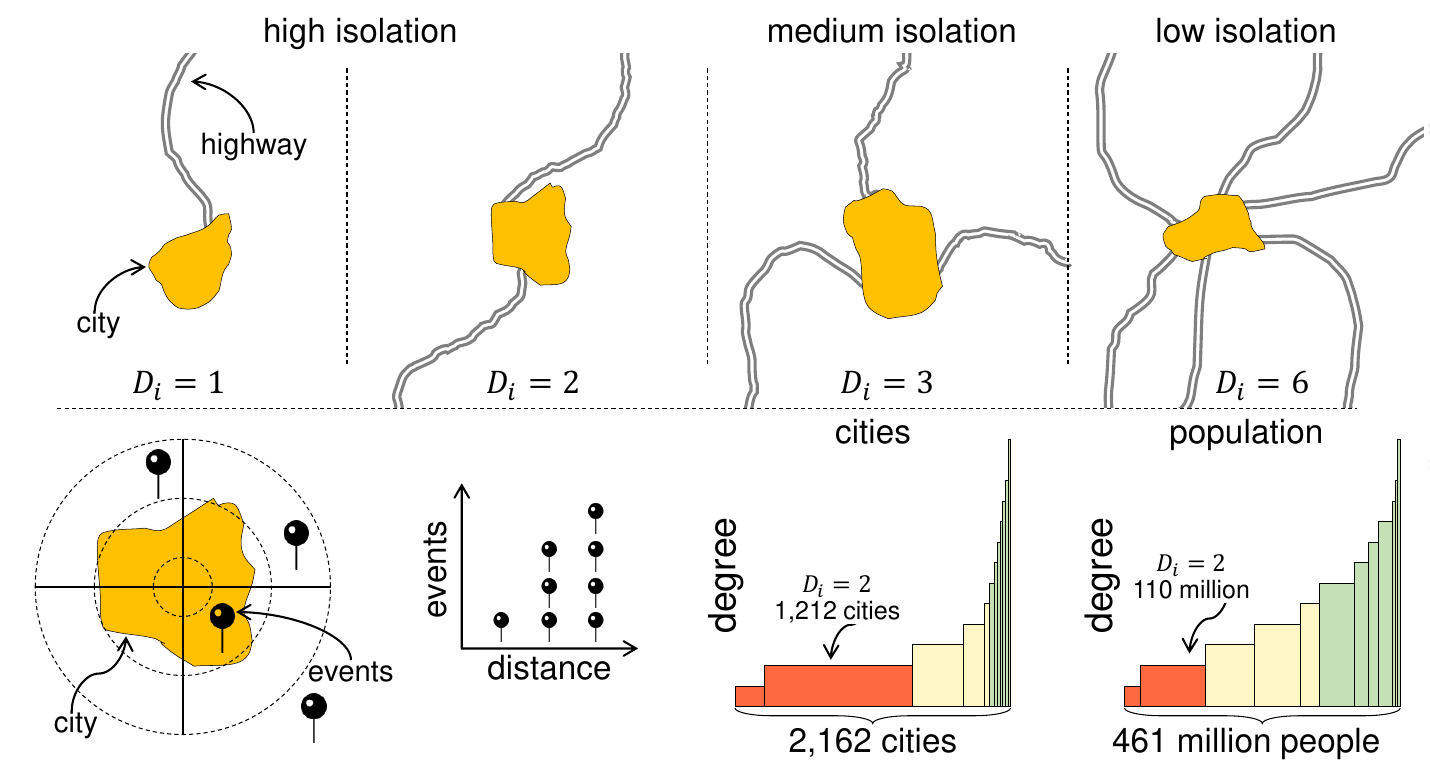}
\caption{\textbf{Urban Isolation and Population Distribution in Africa.} Based on the road network in Africa \cite{prieto2022constructing}, cities are classified by their degree and by their centrality. Most urban areas exhibit a high level of isolation ($D_j \leq 2$), predominantly consisting of smaller cities. Among the 2,162 cities analysed in Africa, 67.1\% are highly isolated, yet these cities comprise only 29.7\% of the considered urban population. Conversely, cities with low isolation constitute merely 3.7\% of the total, but they represent 28.1\% of the urban population.}
\label{FigureDiagramMS}
\end{center}
\end{figure}
}

{
The degree does not capture a city's relationship with the rest of the network but only with its adjacent locations. Thus, a second metric is considered based on the estimated number of journeys that pass through each city. The number of journeys is calculated by assigning trips generated between pairs of cities by a gravity model using the fastest route. The \emph{centrality} of city $j$, expressed as $C_j$, is the number of trips that travel through that city \cite{prieto2022detecting}. It is a form of weighted betweenness which captures the level of centrality of a city based on its size and distance decay effects (details in Section \ref{sec:methods}). Cities with a small centrality $C_j$ are locations with only a few journeys, which tend to be far from big cities and away from significant corridors (additional details in the SM-F). In cities ranking in the lowest 25\% for centrality, lethality rates are 15.1 times higher than those in the top 25\% for centrality, highlighting that remote and peripheral cities exhibit significantly greater lethality compared to their more central counterparts. This disparity extends to violence against civilians, which shows a similar pattern with a ratio of 15.8. Our analysis indicates a clear trend: violence tends to be more concentrated in remote and isolated locations. 
}

{
Notably, violence is more present in small and isolated cities, and it has intensified over the past two decades (Figure \ref{ResultsFigure}-C). Isolation increases the propensity of violence, reduces the capacity of the state to react and allows extremist groups to attack and defend their territory \cite{hammond2018maps}.
}

\section{Discussion} 

{
In African urban landscapes, larger cities report a lower incidence of violence against civilians compared to their smaller counterparts. Despite the higher absolute casualty numbers in larger cities, individuals in smaller cities face a significantly increased risk of politically-motivated violence. This heightened vulnerability extends beyond urban dimensions, particularly affecting small, poorly connected urban centres. These areas frequently become targets for ethnic militias, rebel factions, and Jihadist groups, often outpacing the response capabilities of state forces. The challenge of isolation complicates the situation further, impeding economic growth, elevating the cost of goods and services, diminishing resilience, and restricting access to fundamental services like healthcare and education. Additionally, isolation heightens the susceptibility of cities to politically-motivated violence.
}

{
This work used a static observation of cities (data from Africapolis \cite{Africapolis}) and of the infrastructure that connects them (data from OpenStreetMap \cite{prieto2022constructing, prieto2022detecting}). However, more than 20 years of events from ACLED were used to assess the violence in cities \cite{raleigh2010introducing}. Consequently, results for earlier years must be approached with caution, as cities are rapidly growing, and new agglomerations are also emerging. Nevertheless, in recent years, isolated cities have experienced ten or more times the violence compared to central cities. Hence, even if the observed trend could be influenced by the emergence of new cities in 2021 and 2022, people in isolated cities suffered 14.5 times more casualties related to violence against civilians. Also, there are other data sources to analyse terrorism, such as the Global Terrorism Database (GTD) \cite{GTD2022}. However, GTD does not capture protests and riots (also found here to be sublinear in terms of the number of events and casualties, with details in SM-G). Further, it has been observed that GTD captures roughly half of the terrorist events from ACLED but captures similar patterns and trends  \cite{PrietoTerrorismTimelines}. 
}

{
Events were assigned to cities first by looking at exactly where they were reported and the delineation of urban polygons, and then based on proximity to the centre. Both techniques could be problematic for two reasons. Firstly, the location registered for the events might not be very precise, especially if they occur in remote areas with fewer reference points to accurately capture them. Secondly, small cities have a footprint of only a few km$^2$, whereas major cities occupy larger areas. However, a wide range of distances was tested to match events to cities, and it was found that the results hold, regardless of the length (details in the SM-A to C). Some events do not need to occur precisely inside the polygon of a city for the population to suffer its consequences. Thus, even if the spatial match between events and cities is not perfect, it is more pertinent to detect who suffers violence and why rather than the precision of an event happening exactly inside some delineation of a city or within a few km.
}

{
The dataset might not capture some events, particularly if they were not reported in local media. Although it is possible because the media cannot keep up with reporting minor events in big cities, here we are considering events related to violence against civilians, which cause, on average, 2.5 casualties each. Thus, these events tend to be very visible. In fact, underreporting is more likely in isolated locations, where the media pays less attention \cite{PrietoCEUSMedia}. Incidents in large and central cities, such as Cairo or Lagos, are better documented by the media than in remote locations. Therefore, it is likely that violence against civilians is even more significant in isolated cities than what the data indicates.
}

{
The incidence of violence on the African continent has markedly increased, with isolated cities becoming increasingly vulnerable. From 2010 to 2021, the number of fatalities resulting from violence against civilians escalated dramatically, rising from under 4,000 to nearly 16,000 annually—a 340\% increase, primarily in remote areas. These isolated cities provide strategic advantages for terrorist groups, exploiting the sparse surveillance of limited road networks to establish safe havens. Even basic security measures, such as deploying a few lookouts, can significantly delay state intervention, offering these groups considerable lead time. Addressing this surge in violence necessitates substantial investments in infrastructure and connectivity, along with efforts to strengthen national cohesion.
}

\section{Methods}
\label{sec:methods}

\subsection{Cities and their level of isolation} 

{
Data from Africapolis provides the location and population estimates for all cities considered \cite{Africapolis}. A critical aspect of the analysis of cities is delineating them since defining their boundary is not trivial, and different criteria produce distinct, and often contrasting, results \cite{arcaute2015constructing}. However, in Africapolis, cities were identified using census data and satellite imagery, applying the same definition across the continent. Therefore, it is feasible to analyse data at the continental level. All cities with more than 100,000 inhabitants and smaller towns near main highways are included in the cities dataset. We consider 2,162 cities totalling 460 million inhabitants; roughly half of the continent's population. The network of African highways was constructed from data from OpenStreetMap \cite{OpenStreetMap}. It consists of two datasets: edges and nodes. The 9,159 edges correspond to existing highways on the continent, representing roads connecting cities and road intersections in Africa \cite{prieto2022constructing, prieto2022detecting}. For each edge, an estimate of the time required to travel through that edge is included, aiming to capture the road's straightness, the quality of the road, and other infrastructure attributes. The 7,361 nodes in the network correspond to either cities (2,162) or road crossings (5,199).
}

{
The degree of city \(j\), expressed as \(D_j\), corresponds to the number of highways that connect that city to others. Thus, a city with degree \(D_j = 1\) has only one road to travel from and to that city, whereas a city with degree \(D_j = 2\) is an urban agglomeration growing around a main highway (the highway goes through the city). The degree does not capture a city's relationship with the rest of the network but only with its adjacent nodes. That is, the city \(j\) could have degree \(D_j = 2\), for example, but one of its neighbouring nodes could be another city with degree \(D_k = 1\), thus suggesting that both cities, \(j\) and \(k\), are very isolated, and people will rarely travel to or through them. However, the city \(j\) with \(D_j = 2\) could be adjacent to two large cities on each side, suggesting that it is more central in the network and people are likely to travel more frequently through it. Hence, a second metric trying to capture this nuance is the centrality of a city based on the estimated number of journeys that pass through it. Although one option is to consider the node betweenness directly from the network, it is crucial to consider two aspects of cities. First, city size is highly skewed, meaning that large cities such as Cairo or Lagos are thousands of times bigger than small cities. Second, long-distance journeys are discouraged. Therefore, instead of the node betweenness, a weighted betweenness is considered, aiming to capture the level of centrality of a city. The purpose of this indicator is to estimate the number of journeys travelling through each city. The flow between two cities is estimated using a gravity model. Gravity models are frequently used to analyse spatial interactions, capturing size and distance \cite{GravityModel, Gravity}. In its simplest form, the flow \(F_{o,d}\) between the origin city \(o\) and the destination \(d\) is estimated by
\begin{equation} \label{GravsSimple}
F_{o,d} = \frac{P_o P_d}{N_{o,d}^\gamma},
\end{equation}
where \(P_o\) and \(P_d\) are the population sizes of the origin and destination, respectively, \(N_{o,d}\) represents the travel time, and \(\gamma \geq 0\) captures the impact of travel time on flow. International borders create significant delays and travel frictions (particularly in some parts of Africa \cite{tremolieres2017cross, walther2020mapping}), so here we use an estimate for the travel time across different types of road, and we add two hours for each border crossing, representing the cost that borders impose on the intermediacy of cities and countries \cite{prieto2022detecting}. The value of \(\gamma = 2.8\) has been utilised previously, signalling high travel frictions on the continent \cite{prieto2022detecting}. Subsequently, the estimated flow is assigned to the network through the fastest path. Finally, the centrality is defined as the number of trips that travel through city \(j\), expressed as \(C_j\), and it is estimated by summing the flow that passes through each node in the network when considering all pairs of cities. Formally,
\begin{equation} \label{Interm}
C_j = \sum_{o, d} F_{o,d} H_{o,d}(j),
\end{equation}
where \(H_{o,d}(j) = 1\) if city \(j\) is on the route between \(o\) and \(d\), and zero otherwise. Cities with a small centrality \(C_j\) have few journeys passing through them and, therefore, exhibit high-level isolation. Cities with a high level of isolation tend to have a small degree but also tend to be far from large cities and away from commercial corridors.
}

\subsection{Identifying urban events and comparing lethality across cities}

{
The Armed Conflict Location \& Event Data project (ACLED) is the most comprehensive and detailed database available for analysing violence in Africa \cite{raleigh2010introducing}. Although there are other sources to analyse conflict, the ACLED database stands out as an event-based database, detailing the location and estimated number of fatalities for each event. ACLED categorises six types of events: battles, violence against civilians, explosions/remote violence, strategic developments, riots, and protests. In this analysis, riots and protests are not considered, as they are not typically caused by politically-motivated violent groups (an analysis of protests and riots in the SM-G). From January 2000 to October 2022, ACLED reported more than 182,000 events in Africa associated with politically-motivated violent groups, resulting in more than 560,000 deaths (Table \ref{AcledDataTable}).

\renewcommand{\arraystretch}{0.9}
\begin{table}
\caption{The ACLED dataset has registered more than 180,000 events with more than 560,000 casualties in Africa since the year 2000 related to Battles, Violence against civilians, Explosions and remote violence, and Strategic developments.}
\label{AcledDataTable}
\centering
\resizebox{0.5\textwidth}{!}{%
\begin{tabular}{lrr} 
Type of event & Number of events & Number of casualties \\
 & (2000 - 2022) & (2000 - 2022) \\
\midrule
Battles & 71,323 & 317,795 \\
Violence against civilians & 70,726 & 182,875 \\
Explosions and remote violence &  21,852 & 59,284 \\
Strategic developments & 19,027 & 533 \\
\midrule
Total & 182,928 & 560,487 \\
\end{tabular}
}
\end{table}
}

{
Each event is assigned to a city based on the distance between the event and cities nearby. To assign events to cities, we consider the distance between the location of events and the centre of cities. For event \(i\), we measure the distance in km to the centre of its nearest city \(d_{ij}\) and then assign it to the nearest city \(j\) if \(d_{ij} < \delta\), for some threshold \(\delta > 0\). We will vary the values of \(\delta\) between 1 and 30km to test and ensure that the results do not depend on how events are assigned to cities. The rationale behind this method is that even if violent events do not occur precisely within an urban polygon, they take place within its outskirts, affecting the nearest urban population. Events at a distance larger than \(\delta\) from all cities are classified as rural.
}

{
The impact of events is vastly heterogeneous. Some events are major incidents with hundreds of casualties, whereas other events are much more minor. To compare the lethality across cities, we consider all events that were assigned to the city \(j\) and define the lethality as 
\begin{equation}
\phi_\delta (j) = \frac{\sum_{i \in I_\delta (i, j)} f_i}{P_j},
\label{Eq1}
\end{equation}
where \(P_j\) is the population of city \(j\), and \(I_\delta (i, j)= 1\) if event \(i\) was assigned to city \(j\) and zero otherwise, and \(f_i\) is the number of fatalities of event \(i\). Thus, Equation \ref{Eq1} gives the number of casualties assigned to the city \(j\) divided by its population (with numbers reported in terms of 100,000 inhabitants). The lethality of city \(j\) is a comparable metric across cities that gives the combined impact of politically-motivated violence in each urban area.
}

{
Expressing the number of casualties in city \(i\) as \(L_i(\delta)\) and those related only to violence against civilians as \(V_i(\delta)\), we explore the scaling relations \(L_i \sim P_i^{\beta_L(\delta)}\) and \(V_i \sim P_i^{\beta_V(\delta)}\). A scaling relationship of an indicator \(Y\) with a coefficient \(\beta > 1\) is termed superlinear, indicating that larger cities have disproportionately more of \(Y\) than smaller cities. With \(\beta \approx 1\), the size of a city has little impact on the distribution of \(Y\); with \(\beta < 1\), smaller cities have more of \(Y\), per capita, than larger cities. For instance, for serious crimes in cities in the US, it was observed that \(\beta_S = 1.16 \pm 0.04\), suggesting that larger US cities experience more serious crimes per capita \cite{GrowthBettencourt}. Here, we investigate whether the occurrence of casualties and violence against civilians increases as the city size grows.
}

{
Further, we aim to determine if cities with high isolation exhibit, as a group, greater levels of violence. We construct a metric analogous to the one defined in Equation \ref{Eq1} but applied to a group of cities, such as those identified with high isolation. For a group of cities \(J\), their collective lethality is defined as
\begin{equation}
\phi_\delta^{(J)} = \frac{\sum_{j \in J} \sum_{i \in I_\delta(i,j)} f_i}{\sum_{j \in J} P_j}.
\label{Eq2}
\end{equation}
This equation calculates the total number of casualties across all cities in group \(J\), divided by their combined population size. With \(I\) representing cities with high isolation and \(L\) those with low isolation, we find that \(\phi_{10}^{(I)} = 0.99\) and \(\phi_{10}^{(L)} = 0.13\), indicating that cities with high isolation are eight times more violent than those with low isolation.
}

\subsection{Comparing between cities of different levels of isolation}

{
We define the \emph{isolation impact} \(\theta_\delta\) as the ratio between the lethality of cities with high isolation and those with low isolation. If violence were not associated with the degree of isolation of a city, we would expect to see values of \(\theta_\delta \approx 1\) for certain values of \(\delta\). However, this is not observed, and \(\theta_\delta > 1\) for all values of \(\delta\) (Figure \ref{ResultsFigure}). Similarly, the \emph{centrality impact} is defined as \(\theta_\delta^{c}\) by distinguishing the 25\% of cities with the lowest centrality from the 25\% with the highest centrality and calculating the ratio of their lethality. If centrality did not influence the distribution of violence, \(\theta_\delta^{c} \approx 1\) would be expected; yet, this is not the case.
}

{
Some events result in a high number of fatalities, while most report none. The top 1\% of most lethal events account for nearly 40\% of the casualties documented by ACLED in Africa. A critical examination involves assessing whether the impacts of isolation or centrality stem from just a few high-impact events in a limited number of cities. To test (and reject) this hypothesis, a subset of events is sampled, constructing the same metrics for analysis. Sampling half of the reported events and measuring the impact of isolation and centrality for this subset, the process is repeated 1,000 times to observe potential fluctuations from considering only some events. If the impact of isolation or centrality were due to a few events in a few cities, these events would eventually be excluded, leading to $\theta_\delta^{(s)} \approx 1$ in some iterations. However, across all $\delta$ values and sampling iterations, $\theta_\delta$ and $\theta_\delta^{c}$ significantly exceed one, affirming that isolated cities consistently experience higher violence than central cities.
}

{
Assigning events to cities depends on the parameter \(\delta>0\), considered within the \([1,30]\)km range. For smaller cities, lower \(\delta\) values suffice to encompass their area, whereas larger cities require higher values to include urban and most peri-urban events. With \(\delta = 30\)km, events within this radius are attributed to the city's population. Despite the dependency on \(\delta\), isolated cities consistently exhibit higher lethality across all values of this parameter. Highly isolated cities are found to be at least five times more lethal than central cities, with the ratio rising to seven when focusing solely on violence against civilians (Figure \ref{ScalingResults}). By halving the event sample and recalculating \(\phi_\delta\), the observed lethality ratio for isolated versus central cities remains significant across all \(\delta\) thresholds and sampling iterations (additional details in the SM-D). Thus, the heightened violence in isolated areas is not merely a result of event assignment or the presence of a few high-impact incidents but reflects a broader pattern where isolation, coupled with reduced state presence, elevates violence risk.

\begin{figure}[ht]
  \centering
    \includegraphics[width=.6\textwidth]{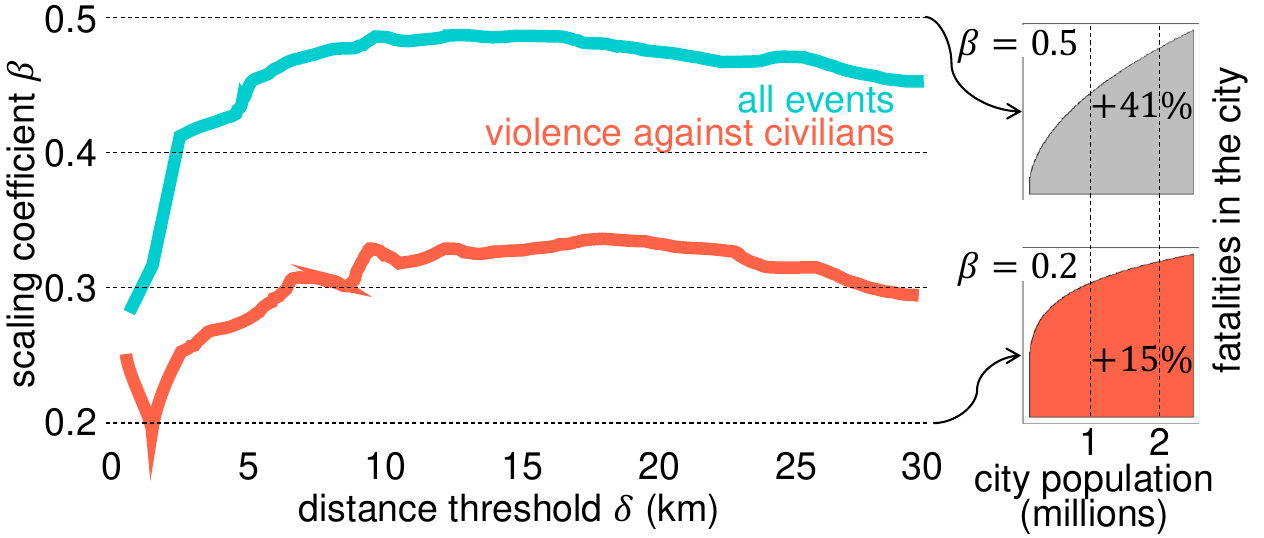}
    \caption{\textbf{Effect of distance on the scaling coefficients.} The scaling coefficient \(\beta_L\) represents the number of casualties from all events, and \(\beta_V\) denotes casualties from violence against civilians (vertical axis), varying with the distance threshold \(\delta\) (horizontal axis) for assigning events to cities. The right side models fatalities based on city size, with two dashed lines indicating casualty numbers for cities of one and two million inhabitants, respectively. Sublinear growth of violence implies that, with \(\beta = 0.3\), a city doubling its population sees only a 23\% increase in fatalities.}
    \label{ScalingResults}
\end{figure}
}

\clearpage

\setcounter{figure}{0}
\setcounter{table}{0}

\renewcommand{\figurename}{Supplementary Figure}
\renewcommand{\tablename}{Supplementary Table}

\section*{Supplementary Materials}

\subsection*{The location of events and the delineation of cities}

{
Detecting which events correspond to an urban agglomeration can be done by computing the intersection between each urban polygon and the coordinates of each event. We take data from the Armed Conflict Location \& Event Data project (ACLED) \cite{raleigh2010introducing}. ACLED is the most comprehensive and detailed database of events happening worldwide. Data from Africapolis is utilised to ascertain the location and size of over 2,000 cities across the continent, employing a uniform definition of an urban area across all countries \cite{Africapolis}. The delineation of each city was constructed by looking at a contiguous built-up surface of not more than 200 m of distance. We identify all events that are recorded within its urban delineation and consider the number of events, $E_i$ and the number of casualties $L_i$ for city $i$. Then, we consider the expression 
\begin{equation}
E_i = \alpha_E P_i^{\beta_E},
\end{equation}
where $\alpha_E$ and $\beta_E$ are two parameters. For each starting year, we compute a Poisson regression to estimate $\alpha_E$ and $\beta_E$ \cite{citeR}. We follow the same strategy for the number of casualties reported within each polygon and obtain $\alpha_L$ and $\beta_L$. Results show that $\beta_E$ and $\beta_L$ have values far from 1, meaning that the number of events and the number of casualties are both sublinear in Africa, regardless of the starting year considered (Table \ref{ScalingBetaBinary}).

\renewcommand{\arraystretch}{0.9}
\begin{table}[ht]
\caption{Scaling coefficients of the number of events depending on city size.}
\label{ScalingBetaBinary}
\centering
\begin{tabular}{l | c c | c c} 
     & \multicolumn{2}{c|}{All events} & \multicolumn{2}{c}{Violence against civilians} \\ 
year & $\beta_E$ & $\beta_F$ & $\beta_E$ & $\beta_F$ \\
\hline
2000 & $0.6926 \pm 0.0011$ & $0.6904 \pm  0.0011$ & $ 0.6422 \pm 0.0023$ & $0.6387 \pm  0.0023$ \\
2001 & $0.6951 \pm 0.0011$ & $0.6878 \pm  0.0011$ & $ 0.641 \pm 0.0023$ & $0.6406 \pm  0.0023$ \\
2002 & $0.7051 \pm 0.0011$ & $0.6904 \pm  0.0012$ & $ 0.6396 \pm 0.0024$ & $0.6396 \pm  0.0024$ \\
2003 & $0.7047 \pm 0.0012$ & $0.7002 \pm  0.0012$ & $ 0.6321 \pm 0.0024$ & $0.6372 \pm  0.0024$ \\
2004 & $0.7032 \pm 0.0012$ & $0.6985 \pm  0.0012$ & $ 0.6279 \pm 0.0025$ & $0.6271 \pm  0.0025$ \\
\hline
2005 & $0.7033 \pm 0.0012$ & $0.6973 \pm  0.0012$ & $ 0.6231 \pm 0.0026$ & $0.6236 \pm  0.0026$ \\
2006 & $0.7016 \pm 0.0012$ & $0.6975 \pm  0.0012$ & $ 0.6213 \pm 0.0026$ & $0.6197 \pm  0.0026$ \\
2007 & $0.6972 \pm 0.0012$ & $0.6962 \pm  0.0012$ & $ 0.6152 \pm 0.0026$ & $0.6188 \pm  0.0026$ \\
2008 & $0.6919 \pm 0.0012$ & $0.689 \pm  0.0013$ & $ 0.6113 \pm 0.0027$ & $0.6085 \pm  0.0027$ \\
2009 & $0.6897 \pm 0.0013$ & $0.6837 \pm  0.0013$ & $ 0.6065 \pm 0.0027$ & $0.6052 \pm  0.0028$ \\
\hline
2010 & $0.683 \pm 0.0013$ & $0.6826 \pm  0.0013$ & $ 0.5971 \pm 0.0028$ & $0.6038 \pm  0.0028$ \\
2011 & $0.683 \pm 0.0013$ & $0.6736 \pm  0.0013$ & $ 0.6025 \pm 0.0029$ & $0.5924 \pm  0.0029$ \\
2012 & $0.6796 \pm 0.0014$ & $0.672 \pm  0.0014$ & $ 0.5936 \pm 0.003$ & $0.5961 \pm  0.003$ \\
2013 & $0.672 \pm 0.0014$ & $0.6667 \pm  0.0014$ & $ 0.5755 \pm 0.0032$ & $0.5858 \pm  0.0031$ \\
2014 & $0.6715 \pm 0.0015$ & $0.655 \pm  0.0015$ & $ 0.5743 \pm 0.0035$ & $0.5659 \pm  0.0034$ \\
\hline
2015 & $0.6801 \pm 0.0017$ & $0.6514 \pm  0.0017$ & $ 0.5993 \pm 0.0038$ & $0.5626 \pm  0.0037$ \\
2016 & $0.689 \pm 0.0018$ & $0.6609 \pm  0.0018$ & $ 0.6047 \pm 0.0041$ & $0.5933 \pm  0.0041$ \\
2017 & $0.6927 \pm 0.0019$ & $0.6711 \pm  0.002$ & $ 0.6039 \pm 0.0045$ & $0.5988 \pm  0.0044$ \\
2018 & $0.6961 \pm 0.0021$ & $0.6683 \pm  0.0022$ & $ 0.5965 \pm 0.005$ & $0.5919 \pm  0.005$ \\
2019 & $0.696 \pm 0.0023$ & $0.6691 \pm  0.0024$ & $ 0.577 \pm 0.0057$ & $0.5776 \pm  0.0057$ \\
\hline
2020 & $0.7018 \pm 0.0027$ & $0.6474 \pm  0.0029$ & $ 0.5857 \pm 0.0067$ & $0.5417 \pm  0.0067$ \\
2021 & $0.732 \pm 0.0035$ & $0.6331 \pm  0.0039$ & $ 0.6089 \pm 0.009$ & $0.5381 \pm  0.0086$ \\
2022 & $0.7138 \pm 0.0064$ & $0.3221 \pm  0.0103$ & $ 0.6616 \pm 0.0163$ & $0.487 \pm  0.0164$ \\
\end{tabular}
\caption*{Scaling coefficients for the number of events and fatalities considered a different starting year.}
\end{table}
}

{
There are, however, two relevant factors to consider regarding this technique. Firstly, the ACLED data is derived from a wide range of local, national, and international sources in over 75 languages, with efforts to ensure that the most specific possible location and time are recorded. Coordinates are often recorded on strategic locations, natural locations, or neighbourhoods. However, this means that the recorded location of some events might not be precise and are often repeated. In Mogadishu, for example, ACLED has reported more than 1,500 events with more than 4,000 fatalities in the exact same building (27 events in the year 2000 and more than 30 events since the year 2020). The location recorded by ACLED is not a perfect representation of the location where those 1,500 events occurred but only an approximation that does not work well at the micro level. In Nouakchott, for example, 98.9\% of all events registered in the city have been recorded with the exact same location, which corresponds to some building that has no particular relevance in a residential neighbourhood (Figure \ref{AcledAfricapolis}). 
}

{
Secondly, there is also an issue related to the Africapolis urban polygons \cite{Africapolis}.  The delineation was carried out before 2015, often with aerial pictures that were taken many years before. Thus, the delineation might be considerably outdated, particularly in countries with rapid population growth (such as Somalia, which grew roughly 40\% between 2010 and 2020). Urban polygons might be bigger than their Africapolis delineation suggests.
}

{
Thus, since the location of events might not be very precise at the local level and the delineation of cities might be outdated, a binary classification of events might be misleading. This might significantly bias the number of events considered for some cities. In Mogadishu, for example, 739 events were reported outside of the city, but within 800 m from the boundary (Figure \ref{AcledAfricapolis}). This corresponds to more than 8\% of the events reported within the boundaries of Mogadishu. Therefore, if those events were recorded in a slightly different location, or if the boundary of the city was constructed based on a more recent aerial picture, it is likely that most of those events would be considered part of Mogadishu and suffered by its population. 

\begin{figure}[ht]
  \centering
      \includegraphics[width=.75\textwidth]{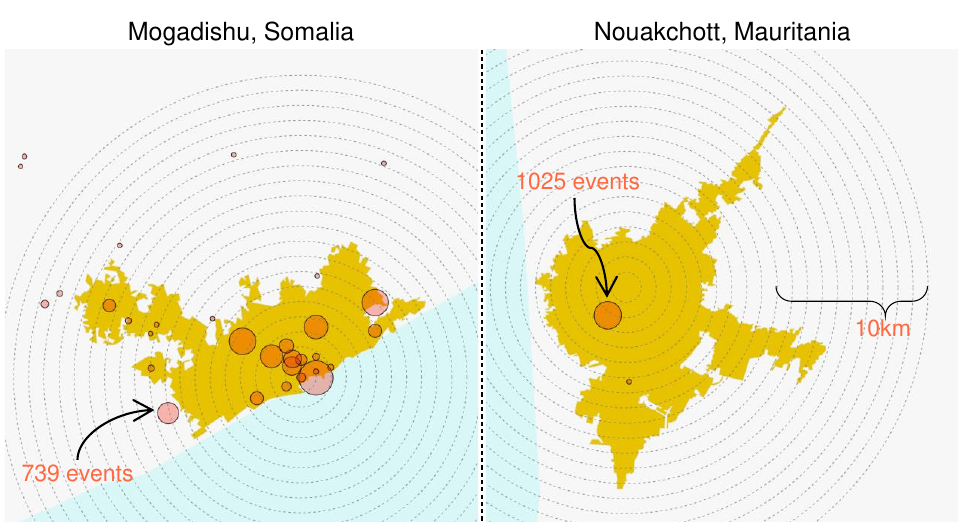}
    \caption{\textbf{Events reported by Acled between 2000 and 2022 in Mogadishu (left) and Nouakchott (right).} The Africapolis urban delineation is the yellow polygon, and the concentric rings are centred in the coordinates of the city centre, and each one has an increment of 1km radius.  }
    \label{AcledAfricapolis}
\end{figure}

Thus, instead of using only the boundary of each city to classify events as being suffered by its population, we also classify events based on their distance to the city centre. 
}

\clearpage

\subsection*{Events depending on the distance threshold}
\label{sup:events}

{
 An event $i$ is attributed to its nearest city $j$ if the distance to its centre $d_{ij} < \delta$, for a given distance threshold $\delta > 0$. For smaller $\delta$ values, fewer events are considered (Figure \ref{DistanceFigureAppendix}). Notably, nearly 2/3 of all events and over half of all casualties occur within 25 km of the cities, despite this area constituting less than 10\% of Africa's surface area. Thus, the majority of events transpire either directly within a city or in its proximity \cite{radil2022urban}.

\begin{figure}[ht]
  \centering
      \includegraphics[width=.5\textwidth]{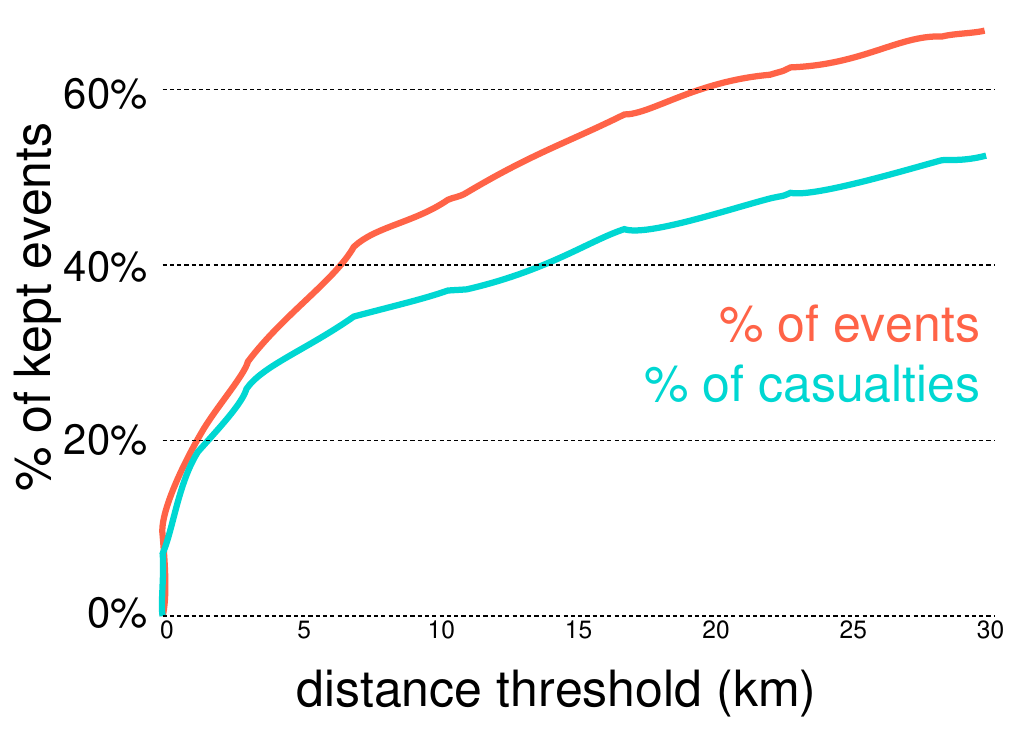}
    \caption{\textbf{Effect of distance on urban rate of events.} Percentage of events and fatalities (vertical axis) within the threshold distance $\delta$ (horizontal axis) of one of the cities.}
    \label{DistanceFigureAppendix}
\end{figure}
}

{
Additionally, data from OpenStreetMap was utilised to digitise the network of all highways across the continent, providing a detailed foundation for analysing the geographic accessibility and connectivity of urban areas \cite{OpenStreetMap, prieto2022detecting, prieto2022constructing}.
}

\clearpage

\subsection*{Reported scaling of violence and crime}
\label{sup:scaling}

{
Reported scaling for different types of crime (Table \ref{ScalingWorld}). The list below consists of an extensive scaling coefficient between different types of violence and city size, including all known values. 

\renewcommand{\arraystretch}{0.9}
\begin{table}
\caption{Observed scaling coefficients for different types of violence in the world.}
\label{ScalingWorld}
\centering 
\begin{tabular}{l l c c} 
Country & Type of crime & $\beta$ & Reference \\
\midrule
Belgium & Burglary & 1.100 & \cite{oliveira2021more} \\
Belgium & Theft & 1.660 & \cite{oliveira2021more} \\
Brazil & Homicides & 1.040 & \cite{ignazzi2014scaling} \\
Brazil  &  Homicides  &  1.150  &  \cite{alves2013distance} \\
Brazil  &  Murders  &  1.150  &  \cite{alves2013scaling} \\
Brazil  &  Homicides  &  1.170  &  \cite{bilal2021scaling} \\
Brazil  &  Homicides  &  1.350  &  \cite{gomez2012statistics} \\
Canada  &  Burglary  &  0.910  &  \cite{oliveira2021more} \\
Canada  &  Theft  &  1.030  &  \cite{oliveira2021more} \\
Colombia  &  Burglary  &  0.940  &  \cite{oliveira2021more} \\
Colombia  &  Homicides  &  1.060  &  \cite{gomez2012statistics} \\
Colombia  &  Theft  &  1.260  &  \cite{oliveira2021more} \\
Denmark  &  Burglary  &  1.150  &  \cite{oliveira2021more} \\
Denmark  &  Theft  &  1.270  &  \cite{oliveira2021more} \\
England and Wales  &  Violence  &  1.120  &  \cite{hanley2016rural} \\
England and Wales  &  Shoplifting  &  1.260  &  \cite{hanley2016rural} \\
England and Wales  &  Bike theft  &  1.273  &  \cite{hanley2016rural} \\
England and Wales  &  Robbery  &  1.550  &  \cite{hanley2016rural} \\
France  &  Theft  &  1.240  &  \cite{oliveira2021more} \\
France  &  Burglary  &  1.290  &  \cite{oliveira2021more} \\
India  &  Murders and homicides  &  0.7788  &  \cite{sahasranaman2019urban} \\
India  &  All crimes  &  0.8699  &  \cite{sahasranaman2019urban} \\
Italy  &  Burglary  &  1.090  &  \cite{oliveira2021more} \\
Italy  &  Theft  &  1.320  &  \cite{oliveira2021more} \\
Latin America  &  Homicides  &  1.100  &  \cite{bilal2021scaling} \\
Mexico  &  Homicides  &  0.970  &  \cite{bilal2021scaling} \\
Mexico  &  Homicides  &  1.120  &  \cite{gomez2012statistics} \\
Mexico  &  Theft  &  1.260  &  \cite{oliveira2021more} \\
South Africa  &  Burglary  &  0.910  &  \cite{oliveira2021more} \\
South Africa  &  Theft  &  0.970  &  \cite{oliveira2021more} \\
Spain  &  Theft  &  1.200  &  \cite{oliveira2021more} \\
UK  &  Theft  &  1.260  &  \cite{oliveira2021more} \\
UK  &  Burglary  &  1.350  &  \cite{oliveira2021more} \\
USA  &  Sworn police officers  &  0.820  &  \cite{banerjee2015competitive} \\
USA  &  Police budget  &  0.880  &  \cite{banerjee2015competitive} \\
USA  &  Requests for police  &  0.960  &  \cite{banerjee2015competitive} \\
USA  &  Burglary  &  0.980  &  \cite{oliveira2021more} \\
USA  &  Burglary  &  1.010  &  \cite{gomez2016explaining} \\
USA  &  Homicides &  1.12  & \cite{bilal2021scaling} \\
USA  &  Theft  &  1.130  &  \cite{oliveira2021more} \\
USA  &  Serious crime  &  1.160  &  \cite{GrowthBettencourt} \\
USA  &  Property crime  &  1.180  &  \cite{chang2019larger} \\
USA  &  All crimes  &  1.260  &  \cite{banerjee2015competitive} \\
USA  &  Robbery  &  1.350  &  \cite{gomez2016explaining} \\
USA  &  Violent crimes  &  1.459  &  \cite{chang2019larger} \\
\end{tabular}
\end{table}
}

\clearpage

\subsection*{Sublinear scaling of violence in African cities}
\label{sup:sublinear}

{
To examine the influence of city size on urban lethality, we define the number of casualties in city $i$ as $L_i(\delta)$ and those specifically from violence against civilians as $V_i(\delta)$. We investigate the relationships $L_i \sim P_i^{\beta_L(\delta)}$ and $V_i \sim P_i^{\beta_V(\delta)}$. Formally, this is expressed as:
\begin{equation} \label{ScalingEq}
L_i = \alpha_L P_i^{\beta_L},
\end{equation}
where $\alpha_L$ and $\beta_L$ are parameters that vary with the distance threshold $\delta$. Given that event and casualty counts are discrete, a Poisson regression is applied to estimate the values of $\alpha_L$ and $\beta_L$.
}

{
The number of fatalities per 100,000 inhabitants is proportional to $P_i^{\beta_L-1}$ once both sides of Equation \ref{ScalingEq} are divided by city size. Thus, a relevant test is to check whether $\beta_L = 1$. We find that for all values of $\delta$ and considering distinct types of events and of starting years, $\hat{\beta_L} < 1$, meaning that small cities have more casualties per 100,000 inhabitants. The same applies to violence against civilians and the obtained values of $\beta_V$. The coefficients obtained for different values of $\delta$ are in Table \ref{TableScaling}.

\renewcommand{\arraystretch}{0.5}
\begin{table}
\caption{Obtained values of $\alpha_L$, $\beta_L$, and of $\alpha_V$ and $\beta_V$ for some values of $\delta$.}
\label{TableScaling}
\centering
\resizebox{0.5\textwidth}{!}{%
\begin{tabular}{ccccc} 
$\delta$ & $\alpha_L$ & $\beta_L$ & $\alpha_V$ & $\beta_V$ \\
\midrule
1 & $3.5954 \pm 0.1486$ & $0.2837 \pm 0.0035$ & $2.7564 \pm 0.2145$ & $0.2540 \pm 0.0064$ \\
5 & $0.5578 \pm 0.1153$ & $0.4402 \pm 0.0017$ & $1.4398 \pm 0.0619$ & $0.2807 \pm 0.0035$ \\
10 & $0.3426 \pm 0.0405$ & $0.4955 \pm 0.0014$ & $0.8308 \pm 0.0297$ & $0.3349 \pm 0.0029$ \\
15 & $0.3798 \pm 0.0062$ & $0.4964 \pm 0.0013$ & $0.9488 \pm 0.0311$ & $0.3339 \pm 0.0026$ \\
20 & $0.46434 \pm 0.0072$ & $0.4865 \pm 0.0012$ & $0.9584 \pm 0.0297$ & $0.3403 \pm 0.0025$ \\
25 & $0.5623 \pm 0.0085$ & $0.4771 \pm 0.0012$ & $1.3283 \pm 0.0397$ & $0.3199 \pm 0.0024$ \\
30 & $0.7401 \pm 0.0108$ & $0.4601 \pm 0.0011$ & $1.8392 \pm 0.0528$ & $0.2999 \pm 0.0024$ \\
\end{tabular}
}
\end{table}
}

{
Trends change between 2000 and 2022. The coefficients obtained by considering only events since a specific year are in Table \ref{TableScalingYear}.

\renewcommand{\arraystretch}{0.85} 
\begin{table}
\caption{Obtained values of $\alpha_L$, $\beta_L$, and of $\alpha_V$ and $\beta_V$ for some starting year and using $\delta = 10$km.}
\label{TableScalingYear}
\centering
\resizebox{0.5\textwidth}{!}{%
\begin{tabular}{ccccc} 
Year & $\alpha_L$ & $\beta_L$ & $\alpha_V$ & $\beta_V$ \\
\midrule
2000 & $0.3308 \pm 0.0059$ & $0.4948 \pm 0.0015$ & $0.7843 \pm 0.0286$ & $0.3369 \pm 0.0029$ \\
2001 & $0.3049 \pm 0.0057$ & $0.4944 \pm 0.0014$ & $0.7823 \pm 0.0288$ & $0.3361 \pm 0.0029$ \\
2002 & $0.2418 \pm 0.0046$ & $0.5070 \pm 0.0015$ & $0.8183 \pm 0.0311$ & $0.3281 \pm 0.0031$ \\
2003 & $0.2471 \pm 0.0048$ & $0.5013 \pm 0.0016$ & $0.9604 \pm 0.0379$ & $0.3107 \pm 0.0032$ \\
2004 & $0.2495 \pm 0.0050$ & $0.4971 \pm 0.0016$ & $0.9551 \pm 0.0390$ & $0.3061 \pm 0.0033$ \\
\addlinespace
2005 & $0.2414 \pm 0.0049$ & $0.4978 \pm 0.0016$ & $0.9924 \pm 0.0411$ & $0.3010 \pm 0.0033$ \\
2006 & $0.2492 \pm 0.0052$ & $0.4935 \pm 0.0017$ &      $1.0129 \pm 0.0424$ & $0.2980 \pm 0.0034$\\
2007 & $0.2717 \pm 0.0058$ & $0.4839 \pm 0.0017$ &      $1.0976 \pm 0.0469$ & $0.2894 \pm 0.0035$\\
2008 & $0.2877 \pm 0.0062$ & $0.4763 \pm 0.0017$ &      $1.1126 \pm 0.0488$ & $0.2850 \pm 0.0035$\\
2009 & $0.2977 \pm 0.0066$ & $0.4690 \pm 0.0018$ &      $1.1616 \pm 0.0516$ & $0.2800 \pm 0.0036$\\
\addlinespace
2010 & $0.3545 \pm 0.0082$ & $0.4508 \pm 0.0019$ &      $1.3324 \pm 0.0607$ & $0.2663 \pm 0.0037$\\ 
2011 & $0.3682 \pm 0.0089$ & $0.4415 \pm 0.0020$ &      $1.2472 \pm 0.0587$ & $0.2667 \pm 0.0038$\\
2012 & $0.3867 \pm 0.0097$ & $0.4325 \pm 0.0020$ &      $1.4692 \pm 0.0716$ & $0.2498 \pm 0.0039$\\
2013 & $0.7320 \pm 0.0204$ & $0.3685 \pm 0.0023$ &      $2.4993 \pm 0.1351$ & $0.1947 \pm 0.0044$\\
2014 & $0.7651 \pm 0.0237$ & $0.3505 \pm 0.0025$ &      $3.0152 \pm 0.1832$ & $0.1648 \pm 0.0049$\\
\addlinespace
2015 & $0.6668 \pm 0.0234$ & $0.3431 \pm 0.0029$ &      $2.1797 \pm 0.1488$ & $0.1739 \pm 0.0055$\\
2016 & $0.6026 \pm 0.0237$ & $0.3349 \pm 0.0032$ &      $2.2268 \pm 0.1665$ & $0.1609 \pm 0.0060$\\
2017 & $0.6141 \pm 0.0273$ & $0.3161 \pm 0.0036$ &      $2.2976 \pm 0.1922$ & $0.1438 \pm 0.0067$\\
2018 & $0.6739 \pm 0.0344$ & $0.2898 \pm 0.0042$ &      $2.9320 \pm 0.2798$ & $0.1097 \pm 0.0077$\\
2019 & $1.2758 \pm 0.0784$ & $0.2158 \pm 0.0050$ &      $5.1565 \pm 0.5751$ & $0.0488 \pm 0.0089$\\
\addlinespace
2020 & $1.2931 \pm 0.1015$ & $0.1851 \pm 0.0064$ &      $5.1951 \pm 0.7130$ & $0.0308 \pm 0.0109$\\
2021 & $0.9820 \pm 0.1211$ & $0.1514 \pm 0.0098$ &      $2.7164 \pm 0.5736$ & $0.0390 \pm 0.0161$\\
\end{tabular}
}
\end{table}
}

\clearpage

\subsection*{Constructing intervals}
\label{sup:intervals}

{
To investigate whether the observed lethality ratio $\phi_\delta$ results from a limited number of high-casualty events, or ``outliers'', in isolated cities, we employ a method where, for a given distance threshold $\delta$, we randomly exclude half of the events and calculate the lethality ratio $\phi_\delta^{(1)}$ based on the remaining events. This procedure is replicated 1,000 times, each with a different subset of events, yielding the set $H_\delta = \{\phi_\delta^{(1)}, \phi_\delta^{(2)}, \dots, \phi_\delta^{(1000)}\}$. We then determine the lower and upper bounds of $H_\delta$. If $\phi_\delta$ were influenced by a few isolated incidents, some iterations of this process would result in $\phi_\delta^{(k)} \approx 1$. Contrary to this hypothesis, our findings do not support such a variation (Figure \ref{DistanceIntervalsAppendix}).
}

{
\begin{figure}[ht]
  \centering
  \includegraphics[width=.7\textwidth]{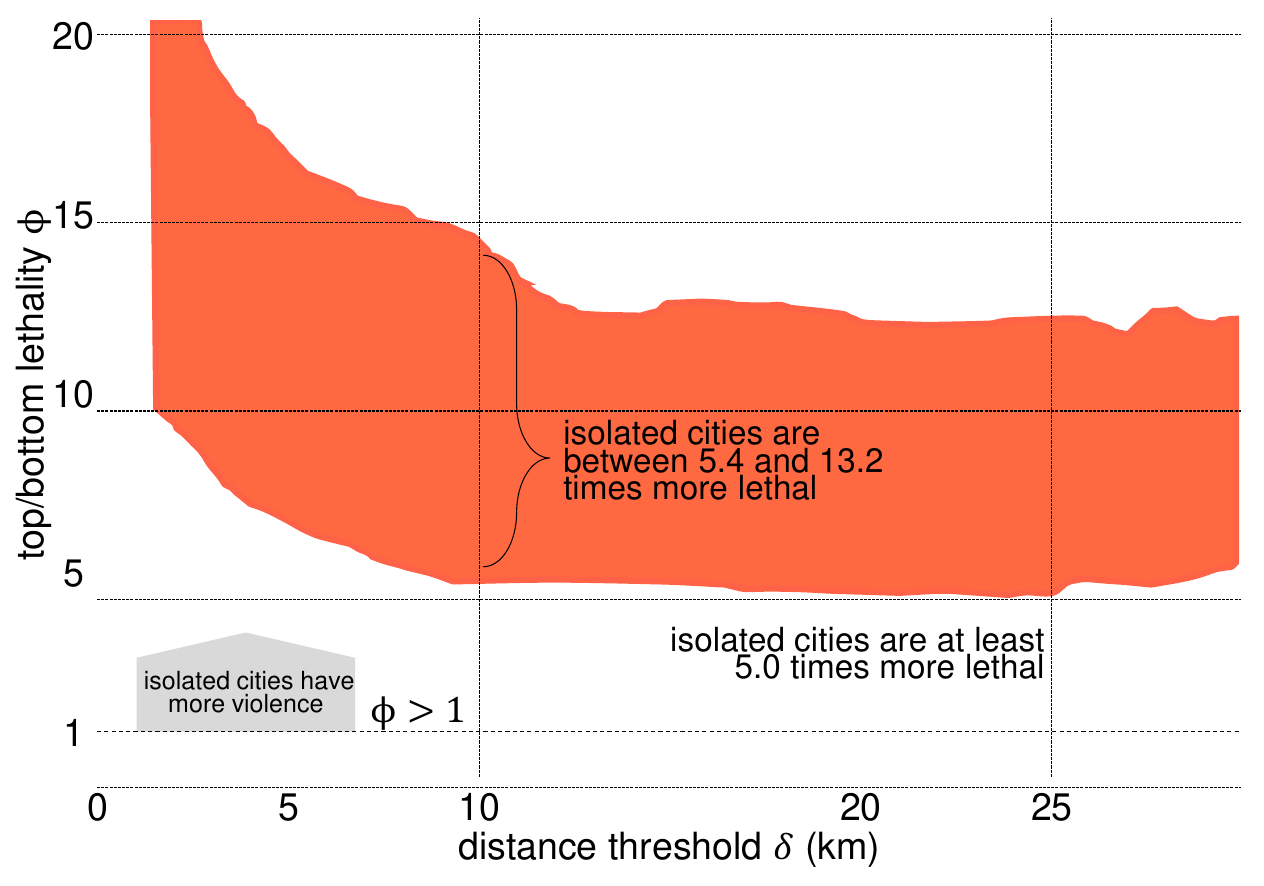}
    \caption{\textbf{Effect of distance on the top to bottom lethality.} Intervals obtained for the lethality ratio $\phi_\delta$ (vertical axis) depending on the distance threshold $\delta$ (horizontal axis).}
    \label{DistanceIntervalsAppendix}
\end{figure}
}

{
For all considered values of $\delta$, it is observed that $\phi_\delta^{(k)} > 5$, indicating that isolated cities are at least five times more lethal compared to central cities. Interestingly, the process of randomly eliminating half of the events often results in an increased ratio of top-to-bottom lethality. This pattern suggests that it is the more central cities that occasionally have a few ``outliers'', while isolated cities exhibit a consistent pattern of violence against their populations, underscoring a systemic issue rather than isolated incidents.
}

\clearpage

\subsection*{Centrality and isolation}

Primary roads, highways and trunks were obtained from OpenStreetMap \cite{OpenStreetMap}, downloaded on March 17, 2021, from http://download.geofabrik.de/ in osm.pbf format. We keep motorways, motorway links, primary roads, primary link roads, trunks and trunk links \cite{prieto2022constructing, prieto2022detecting}. Other road types, including paths, and secondary and tertiary roads, are not considered since they are not frequently used for intraurban mobility. The data gives 5.4 million $x,y$ coordinates as a sequence of vertices of different road segments. Curves are defined with a chain of vertices, so straight roads require fewer vertices than winding roads. We assume that a straight road connects every two consecutive vertices. We keep the physical distance between the two vertices as the length of that segment, using \cite{citeGeosphere, citeR}. For each road, the sum of all the components gives us the road length. In total, we have 415,231km of roads on the continent. We construct a simplified road network of the continent by keeping urban agglomerations with more than 100,000 inhabitants, obtained from Africapolis \cite{Africapolis}, as well as other city attributes, such as its name, country and population in 2015. Each city is then assigned to its closest road coordinates, measured from the centre of the city. For cities with less than 100,000 inhabitants, we measure the distance from the centre to its closest transport node. If a small city is less than 10km away, we assume that the road passes through that city, so we label the closest transport node as a city node. The set of nodes is composed of 2,162 cities and 5.4 million transport nodes. Edges are added to obtain a connected network. Although the data does not contain those edges, we reason that it is possible to travel (perhaps at a slow speed) between two nearby nodes.

The procedure results in a network with cities, crossings, or terminal vertices as the nodes, as well as the travel length and the type of road for each edge that connects the nodes. The result is a connected network with 7,361 vertices, where most of them (71\%) are road nodes and 9,159 edges. We transform travel lengths into travel times, comparing different types of roads (so that travelling on a highway is faster than travelling through other roads), and because travel time enables us to add an extra cost for crossing an international border. Any edge with its extremes on different countries is a border crossing, so we add 120 min for crossing it. We also add extra travel times when crossing any urban agglomeration. We compose it all into some estimated travel time.

The network is used to measure the degree of a city and the intermediacy, a weighted node betweenness \cite{rodrigue2020geography, freeman1978centrality}. We use an expression of the gravity model to estimate the number of journeys between each pair of cities, considering city size and the travel time between them. We estimate the number of trips between each pair of cities and then assign that flow through the fastest path in the network. We then add the number of journeys that pass through each node in the network when all pairs of cities are considered. The intermediacy, $I_k$, is our estimate of the number of journeys that pass through each city. Small values of $I_k$ suggest that only a few trips pass through that city, and large values indicate that many trips pass through that city, even when they did not start or end there. Thus, a city with large values of $I_k$ has a strategic location in the network that connects distinct clusters or a large city size to compensate.

\subsection*{Protests and riots}
\label{sup:ProtestsRiots}

Nearly 80\% of the protests and riots in Africa between January 2000 and October 2022 occurred within 20 km of the centre of a city. Two of three protests and 56\% of the riots also happened within 10 km of the centre of a city. Thus, both are highly urban events. Both types usually have a high frequency but tend to have fewer casualties. The average number of fatalities during a protest is 0.058 and the average number of fatalities during a riot is 0.586 (against 2.585 casualties for each event labeled as violence against civilians).

In terms of the scaling coefficient concerning city size, we find that for protests after January 2010, the number of protests within 20 km of the centre of a city is sublinear with respect to city size (with $\beta = 0.8095 \pm 0.0024$), and it is also sublinear in terms of the casualties related to protests (with $\beta = 0.9638 \pm 0.0099$). Regarding the number of riots in Africa after January 2010, there is also a sublinear pattern regarding city size (with $\beta = 0.7366 \pm 0.0039$) and similarly regarding the number of casualties related to riots (with $\beta = 0.9057 \pm 0.0054$). Regarding the centrality of cities, we find that the lethality ratio $\phi_{20} = 0.8$ for protests and $\phi_{20} = 1.2$ for riots, suggesting that cities have a roughly similar number of protests and riots regardless of their centrality. 

\clearpage

\section*{Funding}

This research is funded by the Federal Ministry of the Interior of Austria (2022-0.392.231).

\bibliographystyle{unsrt}

\end{document}